\documentclass[10pt,conference]{IEEEtran}
\IEEEoverridecommandlockouts
\usepackage{cite}
\usepackage{amsmath,amssymb,amsfonts}
\usepackage{graphicx}
\usepackage{textcomp}
\usepackage[table]{xcolor}
\usepackage[ruled,linesnumbered]{algorithm2e}
\usepackage{hyperref}
\hypersetup{hidelinks}
\usepackage{multirow}
\usepackage{booktabs}
\usepackage{array}
\usepackage{bm}
\usepackage[normalem]{ulem}

\usepackage{subfigure}

\usepackage{microtype}

\definecolor{SBBlue}{RGB}{36,96,158}
\definecolor{SBBlueBg}{RGB}{239,246,253}
\definecolor{SBBlueBorder}{RGB}{120,162,210}
\definecolor{SBGrayBg}{RGB}{249,249,248}
\definecolor{SBGrayBorder}{RGB}{214,214,214}

\ifCLASSOPTIONcompsoc

\else

\fi

\def\BibTeX{{\rm B\kern-.05em{\sc i\kern-.025em b}\kern-.08em
    T\kern-.1667em\lower.7ex\hbox{E}\kern-.125emX}}

\begin{document}

\title{SWE-Bench 5G: Benchmarking AI Coding Agents on Telecom Network Engineering Tasks}

\author{
  \IEEEauthorblockN{Jiao Chen\IEEEauthorrefmark{1},
  Jianhua Tang\IEEEauthorrefmark{1},
  Xiaotong Yang\IEEEauthorrefmark{2},
  and Zuohong Lv\IEEEauthorrefmark{3}}
\IEEEauthorblockA{\IEEEauthorrefmark{1}Shenzhen Smart City Technology Development Group Company, Ltd.\\
\IEEEauthorrefmark{2}Shien-Ming Wu School of Intelligent Engineering, South China University of Technology\\
\IEEEauthorrefmark{3}China Unicom Group Co., Ltd.}
\thanks{Corresponding author: Jianhua Tang (jtang4@e.ntu.edu.sg). J. Chen (202110190459@mail.scut.edu.cn) and J. Tang are with Shenzhen Smart City Technology Development Group Company, Ltd. X. Yang (202510192907@mail.scut.edu.cn) is with the Shien-Ming Wu School of Intelligent Engineering, South China University of Technology.
Z. Lv (lvzh67@chinaunicom.cn) is with China Unicom Group Co., Ltd.}
}

\maketitle

\begin{abstract}
AI coding agents demonstrate strong performance on general-purpose software benchmarks. However, their ability to handle 5G network engineering tasks remains unexplored. We propose SWE-Bench~5G\footnote{Dataset: \url{https://huggingface.co/datasets/tenderzada/SWEBench5G}.}, the first benchmark designed to investigate whether AI coding agents can resolve real-world bugs in 5G core network software. The benchmark collects task instances from three open-source 5G projects, packages each as a self-contained Docker environment with automated fail-to-pass tests, and provides a dual test strategy tailored to the complex runtime dependencies of telecom code. In addition, for instances whose original issues reference 3GPP specification clauses, we construct concise specification context documents, enabling controlled evaluation of whether domain knowledge improves agent performance. Experiments on four LLMs reveal that all models diagnose bugs at rates exceeding 91\%, yet resolve rates remain between 10\% and 30\%, suggesting that both iterative code editing capability and domain knowledge play important roles. The specification injection experiment further confirms that 3GPP excerpts improve resolve rates on specification-dependent bugs, while the gains on generic defensive checks remain limited, indicating that the effect of domain knowledge is conditional on bug type.
\end{abstract}

\section{Introduction}

Large language models (LLMs) have enabled a new class of AI coding agents that autonomously navigate codebases, diagnose bugs, and generate patches~\cite{jimenez2024swebench, hou2024llmse, bariah2024llmtelecom}. Benchmarks such as SWE-Bench~\cite{jimenez2024swebench} and its extensions have driven rapid progress on general-purpose software, primarily Python~\cite{silva2025finetuning}. However, real-world software engineering extends well beyond web applications and data science tooling. In particular, 5G network engineering, where code correctness is governed by hundreds of normative protocol specifications, remains entirely unexamined.

\begin{figure}[!t]
\centering
\includegraphics[width=\columnwidth]{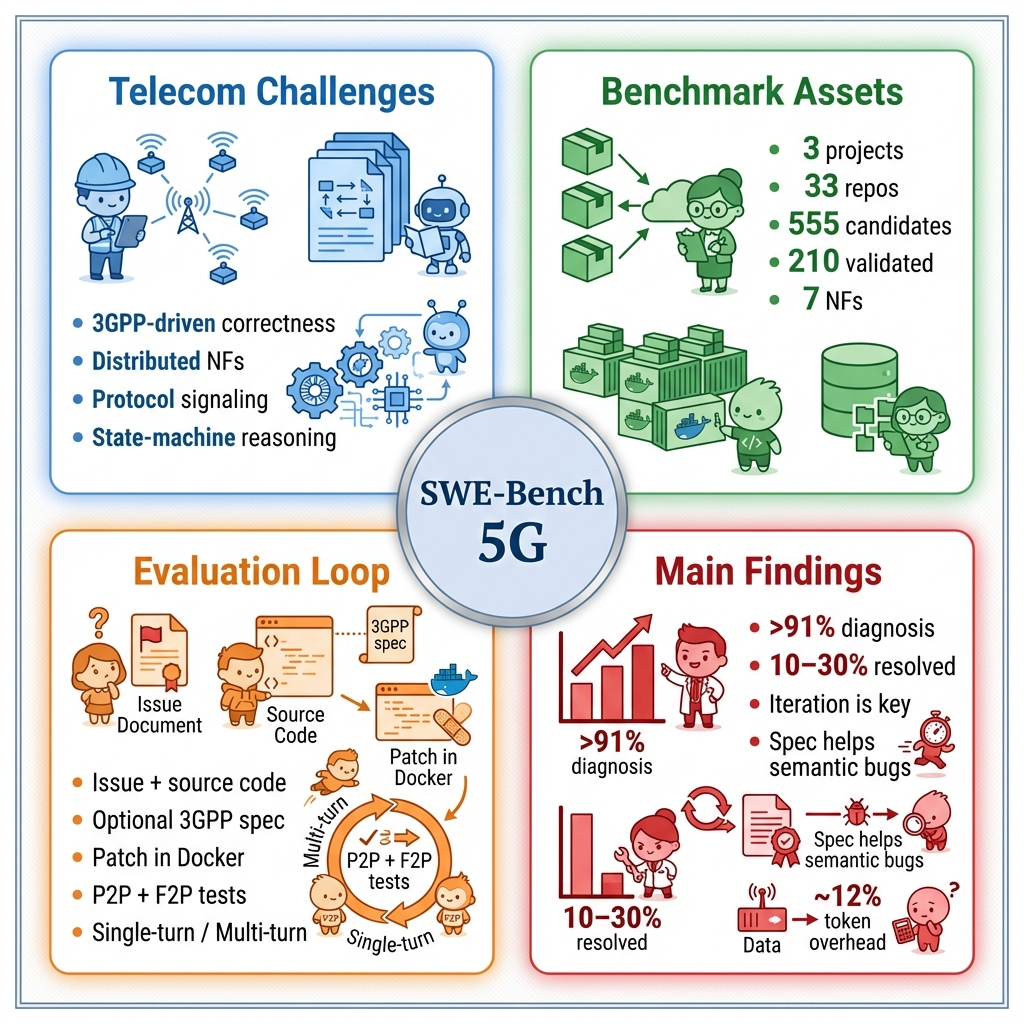}
\caption{Overview of SWE-Bench~5G. The upper-left panel summarizes telecom-specific engineering challenges, the upper-right panel highlights benchmark assets and scale, the lower-left panel shows the evaluation loop, and the lower-right panel presents the main empirical findings.}
\label{fig:unique}
\end{figure}

The 5G core network (5GC) is a representative example of such a domain~\cite{11169296,manias2022nwdaf}. Defined by hundreds of Technical Specifications (TS) from the 3rd Generation Partnership Project (3GPP)~\cite{wang2023road6g, polese2024openran}, 5GC implementations translate precise protocol semantics into production code decomposed across multiple Network Functions (NFs)~\cite{cavalcanti2024tsn5g}. As highlighted in the upper-left panel of Fig.~\ref{fig:unique}, 5G software engineering introduces domain challenges absent from general SWE-Bench settings, including 3GPP-driven correctness, distributed NF interactions, and protocol-state reasoning.

To this end, we propose SWE-Bench~5G, the first benchmark designed to investigate a concrete question: can AI coding agents resolve real-world bugs in 5G core network software? As summarized again in Fig.~\ref{fig:unique}, the benchmark draws task instances from three open-source 5G core implementations, free5GC~\cite{free5gc}, Open5GS~\cite{open5gs}, and Magma~\cite{magma}, spanning 7~NFs; packages each instance as a self-contained Docker environment with fail-to-pass tests; and evaluates both single-turn and multi-turn agent behavior. In addition, since many original issues cite specific 3GPP specification clauses, we leverage these references to construct concise specification context documents, enabling a controlled evaluation of whether providing domain knowledge improves agent performance.

Our contributions are threefold:

\begin{enumerate}
\item A benchmark framework mapping real 5G bugs to the SWE-Bench task format, with a dual test strategy for specification-driven code and Docker-based reproducibility.
\item An evaluation of four LLMs under single-turn and multi-turn agent paradigms, revealing that iterative code editing capability and domain knowledge both contribute to resolving telecom bugs.
\item A controlled experiment on 3GPP specification injection, demonstrating that domain knowledge yields notable gains on specification-dependent bugs while showing limited effect on generic defensive checks.
\end{enumerate}

\section{Related Work}

\subsection{From SWE-Bench to Domain-Specific Benchmarks}

The evaluation of AI coding agents has progressed through several stages. SWE-Bench~\cite{jimenez2024swebench} pioneered this direction by drawing 2,294 task instances from 12 popular Python repositories on GitHub. Each instance pairs an issue description with a fail-to-pass test, enabling automated binary evaluation of whether an agent can produce a correct patch. Early frontier models achieved single-digit resolve rates on this benchmark, catalyzing a wave of agent architecture research~\cite{wang2024pitfalls, fakhoury2024testdriven}.

Subsequent refinements improved evaluation quality. SWE-Bench Lite selected a 300-instance subset for faster iteration, while SWE-Bench Verified introduced human validation to remove ambiguous instances. SWE-Bench Multimodal~\cite{yang2024swebenchmm} extended evaluation to visual software domains, requiring agents to process screenshots and UI specifications alongside code. These efforts remained within the Python ecosystem.

\begin{figure*}[t]
\centering
\includegraphics[width=1.0\textwidth]{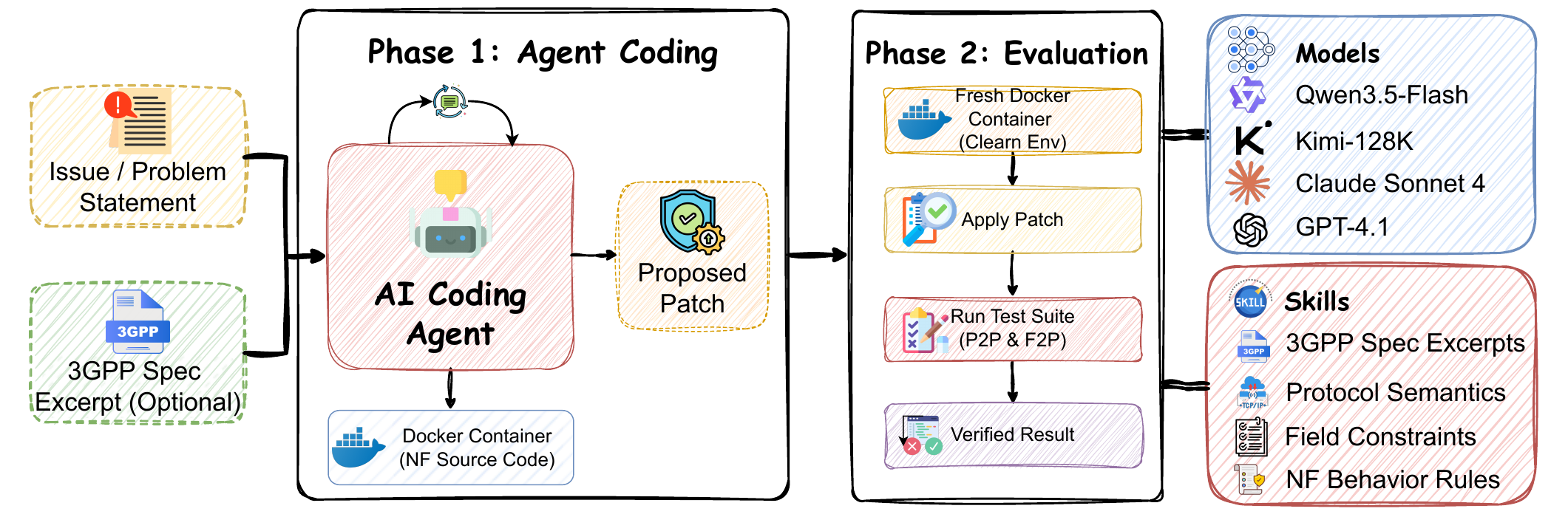}
\caption{SWE-Bench 5G evaluation pipeline. In Phase~1 the agent reads the issue and NF source code, optionally augmented with a 3GPP specification excerpt, and produces a patch. In Phase~2 the patch is applied in a fresh container and validated against pass-to-pass (P2P) and fail-to-pass (F2P) tests.}
\label{fig:overview}
\end{figure*}

\subsection{Expanding Language and Task Scope}

A second generation of benchmarks broadened the language and task coverage. SWE-Bench Mobile~\cite{yao2025swebenchmobile} moved to iOS development with a 500K-line Swift/Objective-C codebase, introducing diff-based intent tests that verify patch correctness through structural source code inspection rather than runtime execution. This addressed the challenge of testing functions with complex runtime dependencies, an approach we adapt for 5G NF functions that require SCTP connections, MongoDB contexts, or inter-NF signaling.

BeyondSWE~\cite{shi2025beyondswe} broadened the resolution scope further to include cross-repository reasoning, domain-specific problem solving, and dependency-driven migrations, packaging each of its 500 instances in a Docker image for full reproducibility. Their experiments revealed that even frontier models plateau below 45\% success on these broader tasks, compared to 80\%+ on SWE-Bench Verified. We adopt the same containerized design and extend it to the telecommunications domain.

\subsection{Skill Injection and Domain Knowledge}

A complementary research direction investigates whether external knowledge can improve agent performance. SWE-Skills-Bench~\cite{han2025sweskills} formalized this question by defining 49 programming skills and evaluating their injection into coding agents. Their central finding is that 80\% of skills provide zero improvement to resolve rates, with utility depending on domain match, abstraction level, and context compatibility. Skills that are too abstract or too specific both fail to help; only skills that closely match the task domain and provide actionable guidance show positive effects.

This finding motivates our investigation of 3GPP specifications as skill documents. Unlike generic coding guidelines, 3GPP Technical Specifications define the authoritative ground truth for correct 5GC behavior and may therefore represent a category of high-utility, domain-matched skill.

\subsection{Gap in Telecommunications}

Despite the growing diversity of benchmarks, all existing work targets general-purpose software. No benchmark evaluates agents on protocol-specified, distributed telecommunications systems. While recent work has examined open-source 5G core implementations~\cite{kousias2023open5g} and their security properties~\cite{giambartolomei2024penetration}, the 5G domain introduces challenges that are absent from Python-centric benchmarks: specification-driven correctness criteria, distributed NF coordination, and domain knowledge underrepresented in LLM training data relative to web development or data science.

\section{Benchmark Design}

\subsection{Overview}

Fig.~\ref{fig:overview} illustrates the SWE-Bench~5G evaluation pipeline. In Phase~1, an AI agent receives an issue description and the NF source code inside a Docker container, optionally augmented with a 3GPP specification excerpt, and produces a patch. In Phase~2, the patch is applied in a fresh container and validated against the test suite.

\subsection{Task Formulation}

Each task instance is a tuple $\mathcal{T} = (I, O, E)$ where $I$ is the input comprising a problem statement with optional 3GPP references and the NF source at the buggy commit, $O$ is the expected output in the form of a unified diff patch, and $E$ is the evaluation consisting of \texttt{PASS\_TO\_PASS} and \texttt{FAIL\_TO\_PASS} test suites. An instance is resolved if and only if all fail-to-pass (F2P) tests pass and all pass-to-pass (P2P) tests remain passing after applying the patch. The complete schema of each task instance is provided in Appendix~A.

\subsection{Source Projects}

We draw task instances from three open-source 5G core network implementations, summarized in Table~\ref{tab:sources}, to improve generalizability and language diversity.

\begin{table}[t]
\centering
\caption{Source projects for SWE-Bench 5G.}
\label{tab:sources}
\begin{tabular}{@{}llccc@{}}
\toprule
Project & Language & Repos & Candidates & Validated \\
\midrule
free5GC~\cite{free5gc}   & Go   & 20 & 320 & 128 \\
Open5GS~\cite{open5gs}   & C    & 1  & 140 & 58  \\
Magma~\cite{magma}        & Go/Python & 12 & 95 & 24 \\
\midrule
Total                     &      & 33 & 555 & 210 \\
\bottomrule
\end{tabular}
\end{table}

free5GC is a Go implementation of the 3GPP Release~15+ 5GC with a modular architecture of one repository per NF. From free5GC alone, validated instances cover seven NFs: Access and Mobility Management Function (AMF), Policy Control Function (PCF), Session Management Function (SMF), Unified Data Management (UDM), NF Repository Function (NRF), Network Slice Selection Function (NSSF), and Authentication Server Function (AUSF). Open5GS is a C implementation widely deployed in research testbeds, with a monolithic repository containing all NFs. Magma, originally developed by Meta, provides a hybrid Go/Python 5G core oriented toward access gateway functionality. The three projects collectively provide language diversity across Go, C, and Python, as well as architectural diversity ranging from modular to monolithic designs.

\subsection{Dual Test Strategy}

Many 5G NF functions cannot be unit-tested directly due to complex runtime dependencies such as Stream Control Transmission Protocol (SCTP) connections, MongoDB sessions, and inter-NF signaling. We therefore employ two complementary test strategies.

Strategy~A, Direct Call, applies when the buggy function accepts simple arguments. We call it directly with crash-triggering inputs and use Go's \texttt{defer/recover} mechanism to detect panics. For example, a PCF policy authorization function can be invoked with a nil \texttt{routeReq} parameter, and the test catches the resulting panic:

\vspace{2pt}
\noindent\texttt{\small defer func()\{ if r:=recover(); r!=nil \{ t.Fatalf("BUG: \%v",r) \} \}()}
\vspace{2pt}

Strategy~B, Diff-Based Intent, inspired by SWE-Bench Mobile~\cite{yao2025swebenchmobile}, verifies that the source code contains the expected fix pattern rather than testing runtime behavior. This forces agents to modify the actual source code while avoiding the construction of complex mock objects. The distribution of strategies across instances is detailed in Table~\ref{tab:dataset}.

As a concrete example, in \texttt{amf\_pr161} the AMF crashes when \texttt{handleUplinkRANConfigurationTransferMain} accesses \texttt{targetRanNodeID.GNbId.GNBValue} without nil-checking \texttt{GNbId}. The handler requires an active SCTP connection, NGAP context, and \texttt{AmfRan} state, making direct invocation impractical. Our Strategy~B test reads \texttt{handler.go} and verifies that the string \texttt{GNbId != nil} appears in the source. On the buggy version this pattern is absent and the test fails; after the agent adds the nil guard the test passes.

\subsection{Validation Criteria}

Each task instance must satisfy a three-step validation before inclusion in the benchmark. First, all existing tests must pass at the parent commit, confirming that the Docker environment is correctly configured. Second, the F2P tests must fail at the parent commit, confirming that the bug is present and detectable. Third, after applying the ground-truth fix via the merged pull request, all tests must pass. Only instances that satisfy all three conditions are included. This process ensures that every instance represents a genuine, reproducible bug with a verified fix.

\subsection{Dataset Construction Pipeline}

Our automated pipeline operates in five stages: (1)~mine closed issues with linked merged pull requests from 33 repositories across the three source projects via the GitHub API, yielding 555 candidates; (2)~filter by quality criteria including clear descriptions and source-modifying PRs; (3)~identify parent (buggy) and fix commits; (4)~construct fail-to-pass tests using Strategy~A or~B; (5)~batch-build Docker images and run three-step validation (existing tests pass, F2P tests fail at parent, all tests pass after fix). Only instances passing all checks are included. Table~\ref{tab:dataset} summarizes the resulting dataset.

\begin{table}[t]
\centering
\caption{SWE-Bench 5G dataset statistics.}
\label{tab:dataset}
\begin{tabular}{@{}ll@{}}
\toprule
Attribute & Value \\
\midrule
Validated instances  & 210 \\
Source projects       & free5GC, Open5GS, Magma \\
Languages            & Go, C, Python \\
NFs covered          & 7 \\
Bug types            & nil/null pointer (89), crash (42), \\
                     & \quad missing validation (35), logic error (27), \\
                     & \quad concurrency (17) \\
Difficulty           & Easy 126, Medium 62, Hard 22 \\
Test strategies      & 68 Direct Call + 142 Diff-Based \\
Avg. tests/instance  & 3.4 \\
3GPP specs referenced & 14 Technical Specifications \\
Docker base images   & \texttt{golang:1.25}, \texttt{gcc:14}, \texttt{python:3.12} \\
\bottomrule
\end{tabular}
\end{table}

\section{Experimental Setup}

\subsection{Agent Paradigms}

We evaluate two agent paradigms. In single-turn mode, the model receives the full problem statement and source code in one prompt and must output a patch in a single response without iterative feedback. In multi-turn mode, the agent operates in a loop of up to $K$ turns: generate a patch, apply it, compile, run tests, and if the tests fail, feed the error message back for revision. We set $K{=}5$.

\subsection{Models}

We evaluate four LLMs selected for diversity across providers, as listed in Table~\ref{tab:models}. All models are accessed through OpenAI-compatible APIs.

\begin{table}[t]
\centering
\caption{Models evaluated in our experiments.}
\label{tab:models}
\begin{tabular}{@{}llll@{}}
\toprule
Model & Provider & API & Paradigm \\
\midrule
Qwen3.5-Flash   & Alibaba   & DashScope  & Single / Multi \\
Kimi-128k        & Moonshot  & Moonshot   & Multi \\
Claude Sonnet 4  & Anthropic & OpenRouter & Multi \\
GPT-4.1          & OpenAI    & OpenRouter & Multi \\
\bottomrule
\end{tabular}
\end{table}

\subsection{Specification-as-Skill Protocol}

In the context of coding agent evaluation, a skill is a structured reference document that encodes domain-specific knowledge, injected into the agent's context at inference time without any model modification~\cite{han2025sweskills}. We adapt this concept to the telecommunications domain by providing 3GPP specification context as skill documents.

Many free5GC and Open5GS issues cite specific 3GPP Technical Specification clauses in the issue description or pull request discussion. For each task instance, we use these references as the starting point to construct a concise specification context document. The document summarizes the relevant protocol semantics, including field optionality constraints and expected NF behavior, and appends a key implication paragraph linking the specification to the task context without revealing the fix. Each resulting document is a self-contained Markdown file averaging 350 tokens. For example, the document for a PCF policy authorization task references TS~29.514, which states that the \texttt{AfRoutReq} field is optional when \texttt{suppFeat} bit~1 is set.

The evaluation follows a paired A/B design. Condition~A provides only the problem statement and source code, while Condition~B additionally appends the specification document to the problem statement. Both conditions use identical Docker images, test suites, and evaluation criteria. We evaluate this on a subset of 50 instances covering both generic bugs, where the fix requires only standard defensive programming, and specification-dependent bugs, where the correct behavior is governed by a specific 3GPP clause.

\section{Results and Analysis}

\subsection{Main Results}

Table~\ref{tab:main_results} presents the resolve rates broken down by bug category. All models are evaluated in multi-turn mode ($K{=}5$) except where noted. We report per-category \%Resolved and an overall average. The best result in each column is shown in bold and the second-best is underlined.

\begin{table*}[t]
\centering
\caption{Main results on SWE-Bench 5G (\%Resolved). Models evaluated in multi-turn mode ($K{=}5$) unless noted. Best in \textbf{bold}, second-best \underline{underlined}. $^\dagger$Single-turn (no feedback loop).}
\label{tab:main_results}
\begin{tabular}{@{}l|ccccc|c@{}}
\toprule
 & NilPtr & Crash & MissValid & LogicErr & Concur. & AVG \\
Model & (89) & (42) & (35) & (27) & (17) & \%Resolved \\
\midrule
Qwen3.5-Flash$^\dagger$ & 0.0 & 0.0 & 0.0 & 0.0 & 0.0 & 0.0 \\
Qwen3.5-Flash & 12.4 & 11.9 & 8.6 & 3.7 & 5.9 & 10.0 \\
Kimi-128k & 11.2 & 9.5 & 11.4 & 7.4 & 5.9 & 10.0 \\
GPT-4.1 & \underline{24.7} & \underline{21.4} & \underline{14.3} & \underline{14.8} & \underline{11.8} & \underline{20.0} \\
Claude Sonnet 4 & \textbf{38.2} & \textbf{28.6} & \textbf{25.7} & \textbf{18.5} & \textbf{17.6} & \textbf{30.0} \\
\bottomrule
\end{tabular}
\end{table*}

Claude Sonnet~4 achieves the highest average resolve rate at 30.0\%, followed by GPT-4.1 at 20.0\%. Qwen3.5-Flash and Kimi-128k each resolve approximately 10\% of instances in multi-turn mode. In single-turn mode ($^\dagger$), Qwen3.5-Flash resolves zero instances despite correctly diagnosing bugs in over 90\% of cases, illustrating the critical role of iterative feedback. Patch application rates (edits successfully written to source, regardless of test outcome) reach 79.5\% for Claude Sonnet~4, 67.1\% for GPT-4.1, 57.1\% for Qwen3.5-Flash, and 51.4\% for Kimi-128k, indicating that the gap between applying a patch and passing tests remains substantial.

Performance varies across bug categories. Nil/null pointer bugs yield the highest resolve rates across all models, as these often require inserting a single guard check. Kimi-128k notably outperforms Qwen3.5-Flash on missing validation bugs, achieving 11.4\% compared to 8.6\%, suggesting that its longer context window helps with specification-dependent fixes. Concurrency bugs are the most challenging category, with even Claude Sonnet~4 resolving only 17.6\%.

\subsection{Single-Turn versus Multi-Turn Analysis}

\begin{figure}[t]
\centering
\includegraphics[width=\linewidth]{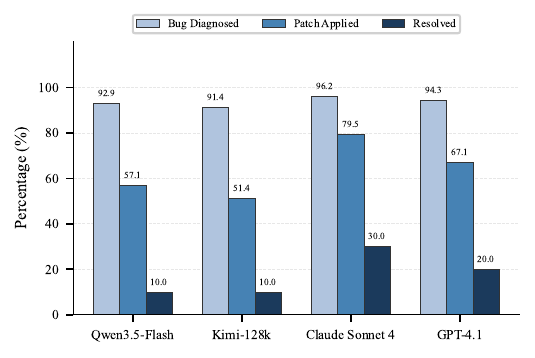}
\caption{Multi-turn evaluation results ($K{=}5$). All models diagnose bugs at rates above 91\% but resolve rates range from 10\% to 30\%, revealing a persistent gap between comprehension and code editing.}
\label{fig:bar_results}
\end{figure}

Fig.~\ref{fig:bar_results} shows the evaluation results across all four models in multi-turn mode. The three-stage funnel, from Bug Diagnosed through Patch Applied to Resolved, is visible for every model. All models diagnose bugs at rates above 91\%, but resolve rates remain between 10\% and 30\%, revealing a persistent gap between bug comprehension and precise code editing.

For Qwen3.5-Flash, the multi-turn loop improves patch application from 34.3\% in single-turn mode to 57.1\% and enables successful resolution of 10.0\% of instances. The feedback mechanism works as intended: when the initial patch causes a compilation error, the error message is fed back, prompting a corrected version in subsequent turns. This result indicates that agentic iteration is a critical factor for telecom software engineering, complementing the domain knowledge required for specification-dependent bugs.

The 30\% resolve rate of Claude Sonnet~4 on SWE-Bench~5G is notably lower than the rates reported for similar models on general-purpose benchmarks. Contributing factors include the stricter type system of Go and C compared to Python, deeply nested NF function signatures, and the exact-match requirement of our SEARCH/REPLACE patch mechanism.

\subsection{Failure Mode Analysis}

We categorize failures into four stages (Table~\ref{tab:failure}). Each instance-model pair is classified by the earliest stage at which the agent fails.

\begin{table}[t]
\centering
\caption{Failure mode breakdown (multi-turn, $K{=}5$). Each cell shows the number of instances whose earliest failure occurs at that stage.}
\label{tab:failure}
\begin{tabular}{@{}lcccc@{}}
\toprule
Failure Stage & Qwen & Kimi & Claude & GPT \\
\midrule
Bug not diagnosed     & 15 & 18 &  8 & 12 \\
Patch format error    & 75 & 84 & 35 & 57 \\
Compile failure       & 20 & 15 & 24 & 22 \\
Incomplete fix        & 79 & 72 & 80 & 77 \\
\midrule
Resolved              & 21 & 21 & 63 & 42 \\
\bottomrule
\end{tabular}
\end{table}

The dominant failure mode across all models is incomplete fix, accounting for 72 to 80 instances per model, where the agent addresses part of the bug but misses additional dereference sites or edge cases. Patch format error is the most prominent bottleneck for Qwen and Kimi, at 75 and 84 instances respectively, where the model generates a correct fix idea but produces a SEARCH block that does not exactly match the source whitespace or formatting. Claude and GPT show substantially fewer format errors, at 35 and 57 respectively, suggesting stronger adherence to structured output formats. This distribution indicates that improving patch application robustness through fuzzy matching could disproportionately benefit weaker models without requiring improvements to the underlying LLM.

\subsection{Specification-as-Skill Results}

Table~\ref{tab:ab_results} presents the specification-as-skill results. We evaluate Claude Sonnet~4, the best-performing model, on a subset of 50 instances, grouped by the 3GPP specification used as the skill document. Pass$^+$ and Pass$^-$ denote resolve rates with and without specification injection, respectively. $\Delta P$ is the skill utility delta, $C^+$ is the average token cost with specification, and $\rho$ is the token overhead ratio.

\begin{table}[t]
\centering
\caption{Specification injection A/B results (Claude Sonnet 4).}
\label{tab:ab_results}
\setlength{\tabcolsep}{3pt}
\begin{tabular}{@{}lrccrrc@{}}
\toprule
Skill Domain & \#Tasks & Pass$^+$ & Pass$^-$ & $\Delta P$ & $C^+$ & $\rho$ \\
\midrule
\multicolumn{7}{@{}l}{Generic bug type (nil/crash)} \\
\midrule
NAS\textsuperscript{a} protocol & 8 & 37.5 & 37.5 & \textcolor{gray}{0.0} & 3.4K & +11\% \\
RAN\textsuperscript{b} signaling & 5 & 40.0 & 40.0 & \textcolor{gray}{0.0} & 3.2K & +9\% \\
NF discovery        & 4 & 25.0 & 25.0 & \textcolor{gray}{0.0} & 3.5K & +13\% \\
Authentication      & 4 & 25.0 & 25.0 & \textcolor{gray}{0.0} & 3.1K & +8\% \\
Slice selection      & 4 & 25.0 & 25.0 & \textcolor{gray}{0.0} & 3.3K & +10\% \\
Mobility mgmt.      & 5 & 40.0 & 40.0 & \textcolor{gray}{0.0} & 3.6K & +14\% \\
\midrule
\multicolumn{7}{@{}l}{Spec-dependent bug type (validation/logic)} \\
\midrule
Policy authorization & 6 & 33.3 & 16.7 & \textcolor{green!60!black}{+16.7} & 3.8K & +15\% \\
Policy control       & 4 & 25.0 & 0.0 & \textcolor{green!60!black}{+25.0} & 3.7K & +14\% \\
Session mgmt.        & 5 & 20.0 & 0.0 & \textcolor{green!60!black}{+20.0} & 3.9K & +16\% \\
System architecture  & 5 & 20.0 & 20.0 & \textcolor{gray}{0.0} & 4.1K & +18\% \\
\midrule
Overall & 50 & 30.0 & 24.0 & \textcolor{green!60!black}{+6.0} & 3.5K & +12\% \\
\bottomrule
\multicolumn{7}{@{}l}{\scriptsize \textsuperscript{a}Non-Access Stratum. \textsuperscript{b}Radio Access Network.}
\end{tabular}
\end{table}

Three skill domains show positive utility deltas: policy authorization, policy control, and session management. These deltas range from +16.7\% to +25.0\% and correspond to bugs where the correct behavior is defined by field optionality or value range constraints in the specification. The remaining seven specifications show zero delta, covering bugs that require only generic defensive programming such as nil guards.

The average token overhead is 12\%, reflecting the concise nature of curated 3GPP excerpts compared to full specification documents.

These results demonstrate that specification utility is conditional on bug type in the telecommunications domain. 3GPP specifications define the ground truth for correct protocol behavior, making them effective for protocol-semantic bugs, while their contribution to bugs requiring only generic defensive programming appears to be marginal.

\section{Conclusion}

We have introduced SWE-Bench~5G, the first benchmark for evaluating AI coding agents on telecom network engineering tasks. The benchmark draws from three open-source 5G core projects, provides a dual test strategy for specification-driven code, Docker-based reproducibility, and an automated construction pipeline. Our evaluation across four LLMs reveals that multi-turn feedback improves resolve rates from 0\% to 10 to 30\%, while the specification injection experiment demonstrates that domain knowledge provides additional gains on specification-dependent bugs, confirming that both agentic capability and domain knowledge are essential for telecom software engineering.

Constructing this benchmark surfaced challenges specific to the telecom domain, notably the near-zero existing test coverage in open-source 5G projects and the difficulty of identifying correct parent commits when merge histories contain partial fixes. Future work includes expanding the dataset with cross-NF coordination tasks, integrating full agentic tools with native file editing capabilities, and evaluating generalization to additional 5G implementations.

\bibliographystyle{IEEEtran}
\bibliography{references}

\appendices

\section{Dataset Schema}

Table~\ref{tab:schema} describes the fields in each task instance of the SWE-Bench~5G dataset, available at \url{https://huggingface.co/datasets/tenderzada/SWEBench5G}.

\begin{table}[h]
\centering
\caption{SWE-Bench 5G dataset schema.}
\label{tab:schema}
\setlength{\tabcolsep}{3pt}
\renewcommand{\arraystretch}{1.1}
\begin{tabular}{@{} l l p{4.2cm} @{}}
\toprule
Field & Type & Description \\
\midrule
\texttt{instance\_id} & string & Unique task identifier \\
\texttt{task} & string & Task type: SingleNF, CrossNF, Protocol, DataPlane \\
\texttt{nf\_type} & string & Network Function: AMF, SMF, PCF, etc. \\
\texttt{repo} & string & Source repository \\
\texttt{image\_url} & string & Docker image for reproducible environment \\
\texttt{patch} & string & Ground-truth fix as unified diff \\
\texttt{parent\_commit} & string & Buggy commit hash \\
\texttt{problem\_statement} & string & Bug description with 3GPP references \\
\texttt{FAIL\_TO\_PASS} & string & Test names that should fail then pass \\
\texttt{PASS\_TO\_PASS} & string & Test names that should always pass \\
\texttt{spec\_reference} & string & Related 3GPP specification sections \\
\texttt{difficulty} & string & Easy, medium, or hard \\
\texttt{affected\_function} & string & Function containing the bug \\
\texttt{lines\_changed} & int & Number of lines in the fix \\
\texttt{files\_changed} & int & Number of files modified \\
\bottomrule
\end{tabular}
\end{table}

\end{document}